\documentclass[a4paper,11pt]{article}
\usepackage{geometry}
\usepackage[T1]{fontenc}
\usepackage[utf8]{inputenc}
\usepackage[english]{babel}
\usepackage{lmodern}
\usepackage{graphicx,graphics}
\usepackage{color}
\usepackage{url} 
\usepackage{epsf,epsfig}
\usepackage{subcaption}
\usepackage{amsmath,amsfonts,amssymb,slashed}
\usepackage{bbold}
\usepackage[colorlinks=true, urlcolor=blue]{hyperref}
\everymath{\displaystyle}
\usepackage{authblk}

\title{\textbf{Hadronic light-by-light scattering contributions to $(g-2)_\mu$ from axial-vector and tensor mesons in the holographic soft-wall model}}
\author[1]{Pietro Colangelo\thanks{\href{{mailto:pietro.colangelo@ba.infn.it}}{pietro.colangelo@ba.infn.it}}}
\author[1]{Floriana Giannuzzi\thanks{\href{{mailto:floriana.giannuzzi@ba.infn.it}}{floriana.giannuzzi@ba.infn.it}}}
\author[1]{Stefano Nicotri\thanks{\href{mailto:nicotri@infn.it}{nicotri@infn.it}}}
\affil[1]{\small \emph{INFN -- Istituto Nazionale di Fisica Nucleare -- Sezione di Bari} \protect\\ \emph{Via Orabona 4, 70125, Bari, Italy}}
\date{}

\begin{document}
\begin{flushright}{BARI-TH/24-755} \end{flushright}
{\let\newpage\relax\maketitle}
\maketitle

\begin{abstract}
    We compute the light axial-vector and tensor meson two-photon transition form factors in the soft-wall holographic model of QCD in the flavor-symmetric case.
    They are used to evaluate the axial-vector and tensor meson contributions to  the anomalous magnetic moment of the muon via the hadronic light-by-light scattering process.
    As expected, these contributions are smaller than the one from pseudoscalar mesons.
    The result for axial-vector mesons is higher than the value found in other approaches.
\end{abstract}

\section{Introduction}
One of the most interesting observables currently under investigation is the anomalous magnetic moment of the muon, $a_\mu=(g_\mu-2)/2$, which is challenging the Standard Model (SM) of fundamental interactions. 
The puzzle deals with the current tension between the measurements \cite{Muong-2:2023cdq} and a SM prediction \cite{Aoyama:2020ynm}.
The most precise measurement has been recently provided by the Muon $g-2$ experiment at Fermilab \cite{Muong-2:2023cdq}, which has improved the precision of the experimental world average by a factor of 2.
From the theoretical point of view, a comprehensive prediction for the SM value has been presented in the White Paper of the Muon $g-2$ Theory Initiative in 2020 \cite{Aoyama:2020ynm}.
The Fermilab result and the expectation quoted in Ref. \cite{Aoyama:2020ynm} deviate at a level of about $5\sigma$. Considering all available data a smaller discrepancy is found \cite{Muong-2:2023cdq}.

In the SM the largest contribution to $a_\mu$ comes from QED processes, and is very precisely known.
Electroweak corrections have also been precisely determined \cite{Jegerlehner:2009ry}.
Then, two kinds of QCD contributions are involved: the leading one corresponds to the hadronic vacuum polarization (HVP) \cite{Colangelo:2018mtw}, the subleading one is due to hadronic light-by-light scattering (HLbL) \cite{Colangelo:2017fiz}.
New discussions can be found in \cite{Melo-Porras:2023xon}.

QCD contributions are the most interesting to look at, since the theoretical uncertainty is dominated by hadronic effects. 
A tension exists between the HVP contribution computed from the $e^+ e^-\to hadrons$ cross section data, used in \cite{Aoyama:2020ynm}, and the lattice QCD result obtained by the BMW collaboration \cite{Borsanyi:2020mff}.
If the latter value is used to obtain the SM $a_\mu$, the discrepancy with the experimental result is reduced to $1.6\sigma$ \cite{DiLuzio:2021uty}.
Moreover, there is a discrepancy between the measurements of the $e^+e^-\to \pi^+\pi^-$ cross section obtained from the BABAR \cite{BaBar:2012bdw} and KLOE \cite{KLOE-2:2017fda} collaborations. 
In this respect, a recent determination by the CMD-3 collaboration \cite{CMD-3:2023alj} is larger than the previous measurements in the energy range up to $\sqrt{s}\simeq 1.2$ GeV.
The latter value would increase the HVP contribution determined in \cite{Aoyama:2020ynm}, reducing the tension between the experimental value of $a_\mu$ and the SM expectation.
On the other hand, the confirmation of the discrepancy between experimental and theoretical values of $a_\mu$ by forthcoming data and theoretical studies would open exciting perspectives for revealing new physics effects \cite{Crivellin:2020zul}.

The theoretical prediction of the HLbL contribution needs to be improved as well in order to meet the precision of $16\times 10^{-11}$ for $a_\mu$ projected for the Fermilab experiment \cite{Colangelo:2022jxc}. 
The HLbL contribution has been computed in a series of studies summarized in Ref. \cite{Aoyama:2020ynm} using a dispersive approach.
This allows a model-independent evaluation but fails in reproducing short-distance constraints (SDCs) \cite{Melnikov:2003xd}, an issue that has been debated in recent years \cite{Masjuan:2020jsf,Colangelo:2019lpu,Colangelo:2019uex}, so a strategy should be found in order to incorporate them.
The dominant contribution comes from the poles of the light pseudoscalar mesons.
The contributions from other states (scalar, tensor and axial-vector states) are smaller, but need to be computed to improve the theoretical accuracy.

In this paper we compute the axial-vector and tensor meson contributions to the HLbL term in the soft-wall model, a QCD phenomenological holographic approach briefly described in Section 2 \cite{Karch:2006pv}. 
Axial-vector mesons play an essential role since they are expected to solve the SDC puzzle reproducing the correct large-$Q^2$ behaviour of the longitudinal four-point function.
In this respect, holographic computations in the hard-wall model with different mechanisms of breaking chiral symmetry have proven to be useful in providing an analytical tool for computing the relevant four-point functions, showing how the sum over the infinite tower of axial-vector states can give the expected result \cite{Cappiello:2019hwh,Leutgeb:2019gbz}.  
This is discussed in Section 3.

Tensor meson contributions are expected to be smaller than contributions from pseudoscalar exchange. 
However, in view of improving the theoretical precision it is important to correctly estimate this contribution, for which there are only a few theoretical studies. 
We compute it by studying the two-photon transition form factor for the helicity-2 meson component in the holographic soft-wall model, obtaining an analytic expression as a function of the photon virtualities.   
The computation is presented in Section 4, with a few technical details collected in the Appendix \ref{sec:AppActionTensor} and \ref{sec:AppAmuTensor}.

\section{The soft-wall model}
The soft-wall (SW) holographic model of QCD is defined in the $5d$ space with background Anti-de~Sitter (AdS) geometry, the bulk, with line element
\begin{equation}\label{eq:metric}
\mathrm{d}s^2=g_{MN}\mathrm{d}x^M \mathrm{d}x^N=\frac{R^2}{z^2} \left(\eta_{\mu \nu} \mathrm{d}x^\mu \mathrm{d}x^\nu - \mathrm{d}z^2\right)\,.
\end{equation}
$R$ is the radius of curvature of the AdS space and $\eta_{\mu \nu}={\rm diag }(1,-1,-1,-1)$. 
We use Greek letters for Minkowski ($4d$) indices, and capital letters for AdS ($5d$) indices.
The fifth $z$ (bulk) coordinate runs in the range $\varepsilon<z<\infty$ with $\varepsilon$ a small ($\varepsilon\to 0$) positive UV cutoff. 
The defining feature of the model is a background (i.e. nondynamical) dilaton field $\phi(z)$ which only depends on the bulk coordinate $z$ and appears in the Lagrangian as $e^{-\phi(z)}$ \cite{Karch:2006pv}. 
A minimal choice is $\phi(z)=c^2 z^2$, where the dimensionful constant $c$, linked to $\Lambda_{\mathrm{QCD}}$, breaks conformal invariance and is responsible for color confinement. 
Such a choice for $\phi(z)$ allows to recover linear Regge trajectories for the spectra of light vector mesons \cite{Karch:2006pv}, light scalar mesons \cite{Colangelo:2008us}, $0^{++}$ and $0^{--}$ glueballs \cite{Colangelo:2007pt,Bellantuono:2015fia}, and $1^{-+}$ hybrid mesons \cite{Bellantuono:2014lra}. 
According to the AdS/CFT dictionary, the global $U(n_f)_L\times U(n_f)_R$ symmetry of QCD is dual to a local (gauge) symmetry in the $5d$ theory, where the fields dual to the QCD left- and right-handed currents $\bar q_{L/R} \gamma^\mu T^a q_{L/R}$ are the massless 1-forms $B_L(x,z)$ and $B_R(x,z)$\footnote{The relation between the $5d$ mass of a $p$-form and the conformal dimension $\Delta$ of its dual $4d$ operator is $m_5^2 R^2=(\Delta-p)(\Delta+p-4)$, for a spin 2 it is $m_5^2R^2=\Delta(\Delta-4)$ \cite{DHoker:2002nbb}.}. 
Such gauge fields are expressed as $B^M_{\{L,R\}}(x,z)=B^{M,a}_{\{L,R\}}(x,z) T^a$, where $T^a=\lambda^a/2$ ($a=0,...,n_f^2-1$) are the generators of $U(n_f)$, with $T^0=\frac{1}{\sqrt{2n_f}} \mathbb{1}_{n_f}$ ($\mathbb{1}_{n_f}$ is the $n_f \times n_f$ identity matrix), and $\mbox{Tr}(T^a T^b)=\delta^{ab}/2$ for $a,b=0,...,n_f^2-1$. 
Vector and axial-vector fields are defined as $V=(B_L+B_R)/2$ and $A=(B_L-B_R)/2$, respectively. 
The $\bar q_R q_L$ operator is dual to a tachyon field $X(x,z)= X_0(z) e^{i\pi^a(x,z) T^a}$, $\pi^a$ describing the nonet of pseudoscalar mesons and $X_0$ parametrised as $X_0(z)=\sqrt{2} v(z) \mathbb{1}_{n_f}$. 
We consider the case of $n_f=3$ light quarks.
The covariant derivative acts on $X(x,z)$ as
\begin{gather}
D_M X = \partial_M X +i [X,V_M]-i\{X,A_M\}\,.
\end{gather}

\section{Axial-vector meson contribution to $a_\mu^{HLbL}$}
The quadratic action for the axial-vector field $A_M(x,z)$ is \cite{Karch:2006pv}
\begin{equation}\label{eq:YMactionA}
    S_A = \frac{R}{k} \int d^5x\, e^{-c^2z^2}\left(-\frac{1}{4g_5^2z} F^a F^a+\frac{4v(z)^2}{z^3} (\partial \pi^a - A^a)^2\right) \,,
\end{equation} 
where $F_{MN}=\partial_M A_N-\partial_N A_M$.
The prefactors $R/k = N_c/16\pi^2$ and $g_5^2 = 3/4$ are fixed by matching the two-point correlation functions of the vector and scalar quark currents to the perturbative QCD expressions \cite{Colangelo:2007pt}.
We set the AdS radius $R=1$.

The coupling of the axial field to two vector fields, contributing to $(g-2)_\mu$, is provided by the Chern-Simons action \cite{Cappiello:2010uy,Colangelo:2011xk,Colangelo:2023een}
\begin{equation}
    S_{CS}=S_{CS}^L-S_{CS}^R
\end{equation}
where
\begin{equation}
    S_{CS}^{L/R} = \frac{N_c}{24\pi^2} \int \, \mbox{Tr}\left(\mathcal{B} \mathcal{F}^2 - \frac{i}{2} \mathcal{B}^3 \mathcal{F} -\frac{1}{10} \mathcal{B}^5 \right)\,,
\end{equation}
$\mathcal{B}=B_M dx^M$ and $\mathcal{F}=(\partial_A B_B)\, dx^A\wedge dx^B$. 
We emphasize that, differently from Ref. \cite{Colangelo:2023een}, we are considering a flavor-symmetric model, and we are not including terms related to the $U(1)_A$ anomaly.
In the gauge $B_z=0$ and keeping only $VVA$ terms, we find, as in the hard-wall model \cite{Leutgeb:2019gbz},
\begin{eqnarray}\label{eq:CSaction}
    S_{CS} &=&  - \frac{N_c}{6\pi^2} \mbox{Tr}(Q_{em}^2 T^a) \varepsilon^{\mu\beta\gamma\delta} \int d^5x\, (3 (\partial_z V_\beta) (\partial_\gamma V_\delta) A_\mu^a + \partial_z( A_\beta^a (\partial_\gamma V_\delta) V_\mu))\,.
\end{eqnarray}
$V_\mu$ is the electromagnetic field, proportional to the light-quark electromagnetic charge matrix $Q_{em}$ and dual to the electromagnetic current. 

$A_\mu$ can be split in a transverse and a longitudinal component: $A_\mu^a=A^{a\perp}_{\mu} + \partial_\mu \varphi^a$.
The longitudinal component of the axial-vector field $(\varphi^{a})$ mixes with the field $\pi^a$ dual to the pseudoscalar current, and they describe pseudoscalar mesons. 
The equations of motion for the transverse and longitudinal components of the axial-vector field come from the quadratic action \eqref{eq:YMactionA} and in the Fourier space they read:
\begin{equation}\label{eq:Aperp}
    \partial_z\left( \frac{e^{-\phi(z)}}{z} \partial_z A^{a\perp}_{\mu}(z)\right)+q^2 \frac{e^{-\phi(z)}}{z} A^{a\perp}_{\mu}(z)-\frac{8g_5^2 v(z)^2 e^{-\phi(z)}}{z^3} A^{a\perp}_{\mu}(z)=0
\end{equation}
\begin{equation}\label{eq:phi}
    \partial_z\left( \frac{e^{-\phi(z)}}{z} \partial_z \varphi^a(z)  \right)+\frac{8 g_5^2 v(z)^2  e^{-\phi(z)}}{z^3} ( \pi^a(z) -\varphi^a(z) )=0 
\end{equation}
\begin{equation}\label{eq:pi}
    \partial_z\left( \frac{e^{-\phi(z)} v(z)^2}{z^3} \partial_z \pi^a(z)\right)+  \frac{e^{-\phi(z)} v(z)^2}{z^3} q^2 (\pi^a(z) - \varphi^a(z))=0 \,.
 \end{equation}

The eigenvalues $q^2=m_n^2$ for the transverse component are found requiring $A_n^{a\perp}(0)=0$ and $\partial_z A_n^{a\perp}(z)=0$ for $z\to \infty$, and the eigenfunction normalization condition is
\begin{equation}\label{eq:normax}
   \frac{R}{k g_5^2} \int_0^\infty dz\, \frac{e^{-\phi(z)}}{z} A_n^{a\perp}(z)^2=1\,.
\end{equation}
If we use
\begin{equation}
    v(z)=m_q z+\sigma z^3\,,
\end{equation}
with $m_q=3.47$ MeV, $\sigma=0.149$ GeV$^3$ and $c=0.388$ GeV, fixed from the pion mass, the pion decay constant and the $\rho$ mass \cite{Colangelo:2023een}, respectively, we find $m_0=1.679$ GeV for the ground-state mass.
In this flavor-symmetric model, $m_q$ is matched to the up and down quark mass.
The axial-vector decay constant, defined as
\begin{equation}
    \langle 0| \bar q\gamma_\mu\gamma_5 T^a q|A_n\rangle = f_n^a m_n \varepsilon_\mu \,,
\end{equation}
can be computed from the expression
\begin{equation}\label{eq:axialdecayconst}
    f_n^a=\frac{R}{kg_5^2} \frac{1}{m_n} \lim_{z\to 0} \frac{e^{-\phi(z)}}{z} \partial_z A_n^{a\perp}(z)\,.
\end{equation}
For the ground state we find $f_0^a=221$ MeV.
These results are independent of the flavour index $a$.

The contribution to muon $g-2$ from the longitudinal component of the axial-vector field in the soft-wall model has been analysed in \cite{Colangelo:2023een}, where the $\eta'$ meson has been included considering the mixing between pseudoscalar mesons and pseudoscalar glueballs \cite{Giannuzzi:2021euy}.

Let us compute the contribution of axial-vector mesons (transverse component) to the correlation function of four vector currents, the $\Pi^A_{\mu\nu\lambda\sigma}$ tensor. As in Ref. \cite{Leutgeb:2019gbz}, we consider it as the sum of the product of two three-point amplitudes times a propagator  over all intermediate axial-vector states \cite{Pauk:2014rta}:
\begin{eqnarray}
    \Pi^A_{\mu\nu\lambda\sigma}(q_1,q_2,q_3,q_4) &=& M_{\mu\nu\alpha}(q_1,q_2) \frac{i P^{\alpha\beta}(q_3+q_4)}{(q_3+q_4)^2-M_A^2} M_{\lambda\sigma\beta}(q_3,q_4) \nonumber\\
    &+& M_{\mu\sigma\alpha}(q_1,q_4) \frac{i P^{\alpha\beta}(q_1+q_4)}{(q_1+q_4)^2-M_A^2} M_{\nu\lambda\beta}(q_2,q_3) \nonumber\\
    &+& M_{\mu\lambda\alpha}(q_1,q_3) \frac{i P^{\alpha\beta}(q_1+q_3)}{(q_1+q_3)^2-M_A^2} M_{\nu\sigma\beta}(q_2,q_4) \,,
\end{eqnarray}
where $M_A$ is the mass of the axial-vector meson, $P_{\mu\nu}=\eta_{\mu\nu}-\frac{q_\mu q_\nu}{M_A^2}$ is the projector for spin 1 mesons, and $q_1,q_2,q_3,q_4$ are the momenta of the incoming photons.

Following \cite{Leutgeb:2019gbz,Pauk:2014rta}, the amplitude $\gamma^* \gamma^*\to A_n^a$ is written as
\begin{equation}\label{eq:AxialAmp} 
    M_{\mu\nu\alpha} (q_1,q_2)= e^2 \varepsilon_{\rho\nu\alpha\tau} (q_1^2 g_\mu^\rho -q_1^\rho q_{1\mu}) q_2^\tau F_{A\gamma^*\gamma^*}(q_1^2,q_2^2) - e^2 \varepsilon_{\mu\rho\alpha\tau} (q_2^2 g_\nu^\rho -q_2^\rho q_{2\nu}) q_1^\tau   F_{A\gamma^*\gamma^*}(q_2^2,q_1^2) 
\end{equation}
where $F_{A\gamma^*\gamma^*}$ is the two-photon transition form factor (TFF). 
\footnote {In general, for axial-vector mesons  three  form factors appear in the decomposition of $M_{\mu\nu\alpha} (q_1,q_2)$   \cite{Roig:2019reh}, while in the holographic model one obtains the amplitude as  in Eq. \eqref{eq:AxialAmp} involving only two form factors. The origin of this mismatch is currently under investigation.}
In the soft-wall model the three-point function of an axial-vector current and two vector currents is obtained from the Chern-Simons action \eqref{eq:CSaction}. 
The eigenfunctions $A_n^{a\perp}(z)$ vanish both at $z=0$ and for $z\to\infty$, so the last term in Eq. \eqref{eq:CSaction} does not contribute.
Then, the TFF of the axial-vector meson $A_n^{a\perp}$ to two photons is given by \cite{Leutgeb:2019gbz}:
\begin{equation}\label{eq:axialTFF}
    F_{A^a_n\gamma^*\gamma^*}(Q_1^2,Q_2^2)=\frac{N_c}{4\pi^2} \mbox{Tr}(Q_{em}^2T^a)\frac{2}{Q_1^2} \int_0^\infty dz\, A_n^{a\perp}(z) \,\partial_z V(z,Q_1^2)\, V(z,Q_2^2)\,,
\end{equation}
where $Q_i^2=-q_i^2$ and $V(z,Q^2)$ is the bulk-to-boundary propagator (BTBP) of the vector field
\begin{equation}\label{eq:VecBTBP}
    V(z,Q^2) = \frac{Q^2}{4c^2} \, \Gamma\left(\frac{Q^2}{4c^2}\right) \, U\left(\frac{Q^2}{4c^2},0,c^2z^2\right)\,,
\end{equation}
with $U$ the Tricomi confluent hypergeometric function \cite{Karch:2006pv,Colangelo:2011xk}. 
In the model the light quark masses coincide and we do not distinguish among mesons with $a=3,8,0$. 
In the TFF the only difference between the states is in the factor $\mbox{Tr}(Q_{em}^2 T^a)$.
We find that the form factors of the lowest-lying axial-vector mesons with $a=8,0$ to two real  photons are $F_{A^8_0\gamma\gamma}=-0.170 \mbox{ GeV}^{-2}$ and $F_{A^0_0\gamma\gamma}=-0.480 \mbox{ GeV}^{-2}$. 
The amplitude for the decay in two real photons vanishes, satisfying the Landau-Yang theorem.
An equivalent two-photon decay width for an axial-vector meson
to decay in one quasireal longitudinal photon and a real photon can be defined as \cite{Pascalutsa:2012pr}
\begin{equation}\label{eq:width}
    \tilde \Gamma_{A_0^a\to\gamma\gamma} = \lim_{Q_1^2\to 0} \frac{m_0^2}{Q_1^2}\frac{1}{2} \Gamma_{A_0^a\to\gamma^*\gamma} =\frac{\pi \alpha^2  m_0^5}{12} F_{A^a_0\gamma\gamma}^2 \,,
\end{equation}
where $Q_1^2$ is the virtuality of the photon $\gamma^*$.
We find $\tilde\Gamma_{A_0^8\gamma\gamma}=5.4$ keV and $\tilde\Gamma_{A_0^0\gamma\gamma}=43$ keV.
In the hard-wall (HW) model $F_{A^8_0\gamma\gamma}=-0.154  \mbox{ GeV}^{-2}$ and $F_{A^0_0\gamma\gamma}=-0.435  \mbox{ GeV}^{-2}$ have been found \cite{Leutgeb:2019gbz}. 
Experimental data for $f_1(1285)$ and $f_1(1420)$ are: $\tilde\Gamma_{\gamma\gamma}=3.5(8)$ keV for $f_1(1285)$ \cite{L3:2001cyf} and $\tilde\Gamma_{\gamma\gamma}=3.2(9)$ keV for $f_1(1420)$ \cite{L3:2007obw}.

\begin{figure}
    \begin{center}
    \includegraphics[width=9cm]{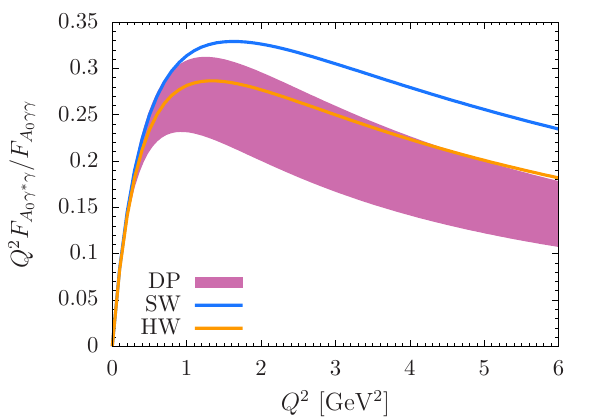}
    \end{center}
    \caption{Axial-vector two-photon transition form factor for one real photon. The blue curve  (SW) shows the result obtained in the soft-wall model; the orange curve  (HW) shows the result obtained in the hard-wall model with bi-fundamental scalar,  denoted as HW1 in \cite{Leutgeb:2019gbz}; the magenta band shows the dipole-parametrization expression (DP) in Eq. \eqref{eq:TFFdipole} including the uncertainty on the parameter $\Lambda$.}
    \label{fig:TFFAxreal}
\end{figure}

\begin{figure}
    \begin{center}
    \includegraphics[width=9cm]{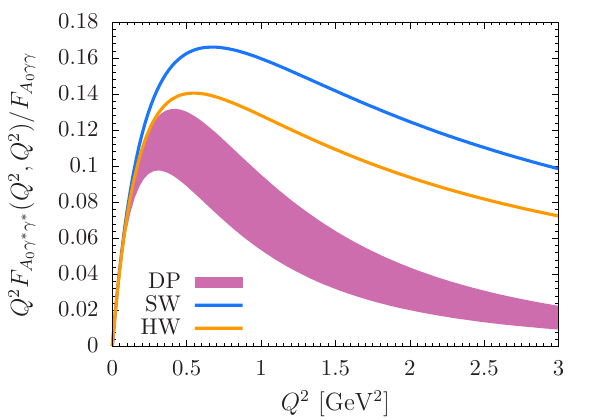}
    \end{center}
    \caption{Axial-vector transition form factor for two virtual photons. Line colors are the same as in Fig. \ref{fig:TFFAxreal}.}
    \label{fig:TFFAx}
\end{figure}

Fig.~\ref{fig:TFFAxreal} shows the $Q^2$ dependence of the axial-vector TFF with one real photon compared to the result in the hard-wall model \cite{Leutgeb:2019gbz} and to the dipole parametrization of Ref. \cite{Pauk:2014rta}
\begin{equation}\label{eq:TFFdipole}
    \frac{F_{A\gamma\gamma}(Q_1^2,Q_2^2)}{F_{A\gamma\gamma}(0,0)}=\frac{1}{(1+Q_1^2/\Lambda^2)^2}\frac{1}{(1+Q_2^2/\Lambda^2)^2}
\end{equation}
with $\Lambda=1040\pm78$ MeV determined from phenomenology for $f_1(1285)$ \cite{L3:2001cyf}.
A similar plot is shown in Fig.~\ref{fig:TFFAx} for $F_{A^a_0\gamma^*\gamma^*}(Q^2,Q^2)$. 
The high-$Q^2$ behaviour in the holographic models is the same as in the dipole parametrization for the TFF with one real photon. 
For two virtual photons the holographic models still find a $Q^{-4}$ decrease of the TFF, while the parametrization in \cite{Pauk:2014rta} produces a $Q^{-8}$ decrease.
As noticed in \cite{Leutgeb:2019gbz,Leutgeb:2021mpu}, defining $Q^2=(Q_1^2+Q_2^2)/2$ and $w=(Q_1^2-Q_2^2)/(Q_1^2+Q_2^2)$, the asymptotic behaviour of the axial-vector TFF for $Q^2\to \infty$ in both the soft-wall and hard-wall models is:
\begin{eqnarray}\label{eq:asyAxial}
    F_{A^a_0\gamma^*\gamma^*}(Q_1^2,Q_2^2) &\xrightarrow[Q^2\to\infty]{}& - N_c \mbox{Tr}(Q_{em}^2T^a) \frac{f_0^a m_0}{Q^4}  \sqrt{1-w} \int_0^\infty dx\, x^4 \, K_0(x \sqrt{1+w}) \,K_1(x \sqrt{1-w}) \nonumber\\
    &=& -  N_c \mbox{Tr}(Q_{em}^2T^a) \frac{f_0^a m_0}{Q^4 w^4}  \left( w (3-2w)+\frac{1}{2} (w+3) (1-w) \log\frac{1-w}{1+w}\right) \,,\nonumber\\
\end{eqnarray}
a result which agrees with the expansion in \cite{Hoferichter:2020lap}.
$K_n(x)$ is the modified Bessel function of the second kind.
The hard-wall and soft-wall models give the same result in this limit since the leading behaviour of vector meson BTBP in both models is $V(z,Q^2) = z \sqrt{Q^2} K_1(z\sqrt{Q^2})$ \cite{Colangelo:2011xk}.
To obtain Eq. \eqref{eq:asyAxial} we have used the expansion of the axial wave function at small $z$ and the definition of the axial-vector decay constant $f_0^a$ in \eqref{eq:axialdecayconst}: 
\begin{equation}
    A_0^{a\perp}(z)  \xrightarrow[z\to 0]{} \left(\frac{R}{k g_5^2}\right)^{-1} \frac{f_0^a m_0}{2} z^2 \,.
\end{equation}

The hadronic light-by-light (HLbL) pole contribution from axial-vector mesons to $(g-2)_\mu$ can be computed from
\begin{equation}
    a_\mu^{HLbL}=\int_0^\infty dQ_1 \int_0^\infty dQ_2 \int_{-1}^1 d\tau\, \rho_a(Q_1,Q_2,\tau)\,,
\end{equation}
where $\rho_{a}$ is defined in the appendix of Ref.~\cite{Leutgeb:2019gbz}.
We find $a^{HLbL}_{\mu,A^3}=9.0\times 10^{-11}$, $a^{HLbL}_{\mu,A^8}=3.0\times 10^{-11}$, $a^{HLbL}_{\mu,A^0}=24\times 10^{-11}$ (the difference is only due to the factor $\mbox{Tr}(Q_{em}^2T^a)$), the sum being $a^{HLbL}_\mu=36\times 10^{-11}$. 
In Table~\ref{tab:amuAx} other determinations are collected, showing that a large uncertainty affects the size of this contribution. 
The soft-wall model, as well as the hard-wall model \cite{Leutgeb:2019gbz}, points toward high values. 
Indeed, it has been noticed \cite{Melnikov:2003xd,Masjuan:2020jsf} that higher $a_\mu$ from axial-vector mesons are found in models able to reproduce the SDCs on the longitudinal four-point function.

   \begin{table}[ht]
    \begin{center}
    \begin{tabular}{ c|cccc } 
    \hline
    Ref. & $a_1^0$ & $f_1(1285)$ & $f_1^\prime(1420)$ & Sum \\
    \hline
    White Paper \cite{Aoyama:2020ynm} & & & & 6 $\pm$ 6\\
    Bijnens \emph{et al.} \cite{Bijnens:1995xf} & & & & 2.5 $\pm$ 1\\
    Hayakawa \emph{et al.}\cite{Hayakawa:1996ki} & & & & 1.738 $\pm$ 0.003\\
    Melnikov \emph{et al.}\cite{Melnikov:2003xd}  & 5.7   & 15.6   & 0.8 & 22 $\pm$ 5\\
    Pauk \emph{et al.} \cite{Pauk:2014rta} &  & 5.0$\pm$2.0 & 1.4$\pm$0.7 & 6.4$\pm$ 2.0\\
    Roig \emph{et al.} \cite{Roig:2019reh} &  0.21$\pm$0.04  & 0.58$\pm$0.11   & 0.015$\pm$0.008  & 0.8$^{+3.5}_{-0.8}$\\
    Leutgeb \emph{et al.} \cite{Leutgeb:2019gbz,Leutgeb:2021mpu} &&&& 29.8-33.2 \\
    Cappiello \emph{et al.} \cite{Cappiello:2019hwh} & 8   & 8  & 12 & 28\\ 
    Masjuan \emph{et al.} \cite{Masjuan:2020jsf} & 5.89 & 10.52 & 1.97 & 18.38\\
    Leutgeb \emph{et al.} \cite{Leutgeb:2022lqw} & 7.1-7.8 & 4.3-5.7 & 13.6-14.3 & 25.0-27.8 \\
    Radzhabov \emph{et al.} \cite{Radzhabov:2023odj} & & & & 3.6$\pm$1.8 \\
    This work & & & & 36\\
    \hline
    \end{tabular}
    \end{center}
    \caption{\small HLbL contribution to $a_\mu$ ($ \times 10^{11}$) of the axial-vector mesons $a_1, f_1, f_1^\prime$ in the indicated references. For more results see Table 13 in \cite{Aoyama:2020ynm}. All values refer to the ground states, with the exception of the one quoted as Ref. \cite{Cappiello:2019hwh}  corresponding to the whole meson tower.  In \cite{Leutgeb:2019gbz} the same model (called HW2) of Ref. \cite{Cappiello:2019hwh} is considered, with flavor-symmetric normalizations, and the ground state contributions are 80\% of the whole meson tower.}
    \label{tab:amuAx}
   \end{table}

Notice that, as stated in Ref.~\cite{Pauk:2014rta}, when calculating $a^{HLbL}_\mu$ the sum over polarizations of spin-1 particles gives the projector $P_{\mu\nu}=\eta_{\mu\nu}-\frac{q_\mu q_\nu}{M_A^2}$, containing both transverse and longitudinal contributions. 
We can separate them as the sum of a transverse $P^\perp_{\mu\nu}=\eta_{\mu\nu}-\frac{q_\mu q_\nu}{q^2}$ and a longitudinal projector $P^{||}_{\mu\nu}=\frac{q_\mu q_\nu}{q^2} \frac{M_A^2-q^2}{M_A^2}$. 
Using the transverse projector, we find $a^{HLbL,\perp}_\mu=16\times 10^{-11}$, and using the longitudinal one we find $a^{HLbL,||}_\mu=20\times 10^{-11}$, both values including the contributions from $a=3,8,0$.

The contributions from the excited states are smaller. 
Even though the soft-wall model is not able to correctly describe these states in the axial-vector sector, we have computed their contributions to $a_\mu^{HLbL}$ with the results collected in the Table \ref{tab:axialexcited}.
Moreover, following \cite{Leutgeb:2019gbz}, we have computed the Green's function of the axial-vector field and used it to determine the contribution to $a_\mu^{HLbL}$ from the whole tower of axial-vector mesons.
We  find $a_\mu^{HLbL}=41.3 (18.1+23.2)\times 10^{-11}$, with the values in parentheses corresponding to the transverse and longitudinal contributions. This value is included also in Table \ref{tab:axialexcited},   last column.
This result agrees with the ones obtained in Ref. \cite{Leutgeb:2021mpu} in the hard-wall model in the flavor symmetric case, where, depending on the different description of the scalar sector and chiral symmetry breaking, the contribution from the whole tower of axial-vector mesons lies in the range $39.9 (17.2+22.7) - 43.3 (18.3+25.0)\times 10^{-11}$.
In Ref. \cite{Leutgeb:2022lqw} different masses for up/down and strange quarks have been considered and a Witten-Veneziano mass has been included, finding that the sum over the sectors for the whole meson tower is in the range $(30.5-33.7)\times 10^{-11}$ depending on how the parameters are set.
\begin{table}[ht]
    \begin{center}
    \begin{tabular}{ c|ccccc|c } 
    \hline
     & $n=0$ & $n=1$ & $n=2$ & $n=3$ &  $n=4$ &  $\sum_n$ \\
    \hline
    Mass (GeV) & 1.7 & 2.7 & 3.5 & 4.2 & 4.9 &\\
    $a_\mu^{HLbL}$ & 36 & -0.5 & 3.7 & -0.2 & 1.3 & 41.3  \\
    (T+L) &  (16+20) & (-0.4-0.1) & (1.8+1.9) & (-0.13-0.07) & (0.65+0.67) & (18.1+23.2) \\
    \hline
    \end{tabular}
    \end{center}
    \caption{\small Mass of the excited axial-vector mesons  in the soft-wall model ($n=0$ corresponds to the ground state), and their contribution to $a_\mu^{HLbL}$ ($ \times 10^{11}$) . The value of $a_\mu^{HLbL}$ in the last column is the contribution from the whole tower of axial-vector mesons computed by the Green's function. In the last row we have emphasized the transverse (T) and longitudinal (L) contributions to $a_\mu^{HLbL}$ ($ \times 10^{11}$).}
    \label{tab:axialexcited}
   \end{table}

Let us conclude this Section with a comment on the SDCs.
The longitudinal component of the HLbL four-point function can be expressed as
\begin{equation}
     \Pi_{\mu\nu\alpha\beta}^{||}(q_1,q_2,q_3,q_4) = T^{(1)}_{\mu\nu\alpha\beta} \bar\Pi_1 + T^{(2)}_{\mu\nu\alpha\beta} \bar\Pi_2 + T^{(3)}_{\mu\nu\alpha\beta} \bar\Pi_3 
\end{equation}
with $T^{(1)}_{\mu\nu\alpha\beta}=\varepsilon_{\mu\nu\rho\sigma} \varepsilon_{\alpha\beta\rho'\sigma'} q_1^\rho q_2^\sigma q_3^{\rho'} q_4^{\sigma'}$, and $T^{(2,3)}$ and $\bar \Pi_{2,3}$ obtained by crossing operations \cite{Cappiello:2019hwh}. 
The SDC conditions are \cite{Melnikov:2003xd}:
\begin{eqnarray}\label{eq:OPEq1neqq3}
     \lim_{Q_3\to\infty} \lim_{Q\to\infty} Q^2 Q_3^2 \bar\Pi_1(Q,Q,Q_3) = -\frac{2}{3\pi^2}
 \end{eqnarray}
 \begin{eqnarray}\label{eq:OPEq1eqq3}
    \lim_{Q\to\infty} Q^4 \bar \Pi_1(Q,Q,Q)= -\frac{4}{9\pi^2}\,.
 \end{eqnarray}
They have been checked in the hard-wall model in Ref. \cite{Leutgeb:2019gbz,Leutgeb:2021mpu} and they are also obtained in the soft-wall model and agree with the considerations of Ref. \cite{Masjuan:2020jsf}. The factorized contribution from pions scales as $1/Q^6$, of higher order in the  $Q^2\to\infty$ expansion than the result from the OPE. The longitudinal contribution from axial-vector mesons (obtained from the projector $\frac{q_\mu q_\nu}{q^2} \frac{M_A^2-q^2}{M_A^2}$) is instead of the same order as the OPE result: it matches the OPE result in the region $Q^2_1 \sim Q^2_2 \gg Q^2_3 \gg m^2_\rho$, while  in the region $Q^2_1 = Q^2_2 = Q^2_3 \gg m^2_\rho$ the correct power behaviour is reproduced although the numerical factor is, both in the hard-wall and soft-wall models, $81\%$ of the value in \eqref{eq:OPEq1eqq3}:
\begin{equation}
     \lim_{Q\to\infty} Q^4 \bar \Pi_1(Q,Q,Q) \sim -\frac{0.36}{\pi^2}\,. 
 \end{equation}

\section{Tensor meson contribution to $a_\mu^{HLbL}$}\label{Sec:Tensor}
Tensor mesons with $J^{PC}=2^{++}$ can be described in holographic models by a field dual to the $j_{\mu\nu}=\bar q \frac{1}{2} \left(\gamma_\mu i\overleftrightarrow{D}_\nu + \gamma_\nu i\overleftrightarrow{D}_\mu \right)q$ operator, i.e. by a symmetric tensor. 
In \cite{Katz:2005ir,Mamedov:2023sns} the spin-2 field $h_{\mu\nu}$ (graviton) has been introduced as the 4$d$ fluctuation of the metric \eqref{eq:metric}:
\begin{equation}\label{eq:metricfluct}
    ds^2=g_{MN}dx^M dx^N =\frac{1}{z^2} (\eta_{\mu\nu}+h_{\mu\nu}) dx^\mu dx^\nu-\frac{1}{z^2} dz^2\,.
\end{equation}
The action for $h_{\mu\nu}$ is obtained from the Einstein-Hilbert action: 
\begin{equation}
    S_{EH} = -2k_T\int d^5x \,\sqrt{g} \, (\mathcal{R}+2\Lambda)\,,
\end{equation}
where $\mathcal{R}$ is the Ricci scalar and $\Lambda$ the cosmological constant.
In the soft-wall model,  in the quadratic approximation the action for the transverse traceless field $h_{\mu\nu}$ is (see the Appendix \ref{sec:AppActionTensor}):
\begin{equation}\label{eq:actionTensor}
    S = -\frac{k_T}{2} \int d^5x \, \frac{e^{-\phi}}{z^3} \eta^{\mu\alpha} \eta^{\nu\beta} (\partial_z h_{\mu\nu} \partial_z h_{\alpha\beta} + h_{\mu\nu} \Box h_{\alpha\beta})\,.
\end{equation}
The coefficient $k_T$ is fixed from the two-point correlation function.
The field $h_{\mu\nu}$ has helicity $\pm2$, and is a flavor singlet (representing singlet tensor mesons).

Higher spin mesons have been studied in \cite{Germani:2004jf,Karch:2006pv,dePaula:2008fp,Abidin:2008ku,Gutsche:2011vb,Lyubovitskij:2023lrp} in holographic models, and a review on the description of fields with  arbitrary spin in light-front holographic QCD can be found in \cite{Brodsky:2014yha}.  
In \cite{Karch:2006pv} a rank-2 symmetric tensor field $H_{AB}$ is introduced, and, in the axial gauge $H_{Az}=0$, the field $h_{\mu\nu}$, defined from $H_{\mu\nu}=h_{\mu\nu}/z^2$, obeys the same action \eqref{eq:actionTensor}.

From the action \eqref{eq:actionTensor}, the equation of motion in the Fourier space is
\begin{equation}\label{eq:eomTensor}
    \partial_z\left(\frac{e^{-\phi}}{z^3} \partial_z h_{\mu\nu}\right) +\frac{e^{-\phi}}{z^3} q^2 h_{\mu\nu}=0\,. 
\end{equation}

The two-point correlation function is found from the on-shell action:
\begin{equation}
    S_{os}=\lim_{z\to 0} \frac{k_T}{2}\int d^4x \frac{e^{-\phi}}{z^3} \eta^{\mu\alpha} \eta^{\nu\beta} h_{\mu\nu} \partial_z h_{\alpha\beta}
\end{equation}
deriving it with respect to the sources of the $4d$ tensor operator \cite{Witten:1998qj}.
In the Fourier space the field $h_{\mu\nu}$ is related to the source $\hat h_{\mu\nu}$ by: $h_{\mu\nu}(z,q^2)=h_B(z,q^2)\hat h_{\mu\nu}(q^2)$, where $h_B$ is the bulk-to-boundary propagator
\begin{equation}
    h_B(z,q^2)=\Gamma\left(2-\frac{q^2}{4c^2}\right) U\left(-\frac{q^2}{4c^2},-1,c^2z^2\right)\,
\end{equation}
obtained solving the equation of motion \eqref{eq:eomTensor} with boundary conditions $h_B(0,q^2)=1$ and finiteness of the action.
The two-point function reads
\begin{equation}
    \Pi^{\mu\nu\rho\sigma}=\frac{\delta^2 S_{os}}{\delta \hat h_{\mu\nu}\delta \hat h_{\rho\sigma}} = k_T P^{\mu\nu\rho\sigma}\lim_{z\to 0} \frac{e^{-\phi}}{z^3} h_B \partial_z h_B\,,
\end{equation}
where $P_{\mu\nu\rho\sigma}=\frac{1}{2}(\eta_{\mu\rho}\eta_{\nu\sigma}+\eta_{\mu\sigma}\eta_{\nu\rho}-\frac{2}{3}\eta_{\mu\nu}\eta_{\rho\sigma})$ is the transverse projector.
For $Q^2=-q^2\to\infty$ one has ($\nu$ is an energy scale)
\begin{equation}
    \Pi^{\mu\nu\rho\sigma} \to -k_T P^{\mu\nu\rho\sigma} \frac{Q^4}{8} \log (Q^2/\nu^2)
\end{equation}
to be compared with \cite{Katz:2005ir,Novikov:1981xi}
\begin{equation}
    \Pi^{\mu\nu\rho\sigma}_{QCD} \to -P^{\mu\nu\rho\sigma} \left(\frac{N_c N_f}{160\pi^2}+\frac{N_c^2-1}{80\pi^2}\right) Q^4 \log (Q^2/\nu^2)\,.
\end{equation}
This condition fixes $k_T=\left(\frac{N_c N_f}{20\pi^2}+\frac{N_c^2-1}{10\pi^2}\right)$.

Eigenvalues and eigenfunctions are found by solving Eq. \eqref{eq:eomTensor} requiring $h_n(0)=0$ and normalisation
\begin{equation}
    \int_0^\infty dz\, \frac{e^{-\phi(z)}}{z^3} h_n(z)^2=1\,.
 \end{equation}
The spectrum is $m_n^2=4 c^2(n+2)$ and the wave functions are  \cite{Mamedov:2023sns}
 \begin{equation}\label{eq:TensorWF}
     h_n(z)=\sqrt{\frac{2 }{(n+1)(n+2)}} \, c^3 z^{4} \, L_n^2(c^2z^2)\,,
 \end{equation}
in terms of   the generalized Laguerre polynomials $L_n^\lambda(z)$.
The residues of the poles of the two-point function are $R_n=k_T 8c^6 (n+1) (n+2)$. From $R_n=f_n^2 m_n^4$ we find the decay constants $f_n = c\sqrt{k_T (n+1)/(2(n+2))}$.
For $c=0.388$ GeV we obtain $m_{f_2}=1.097$ GeV and $f_{f_2}=69$ MeV (corresponding to $n=0$), while the first excitation ($n=1$) has mass $m_{1}=1.344$ GeV and decay constant $f_1=80$ MeV.
The decay constant $g_f=f_{f_2}/m_{f_2}= \sqrt{5/2}/(8 \pi)=0.063$ can be compared to Ref. \cite{Aliev:1981ju} where $g_f=0.040$ is found, while in the hard-wall model one has $g_f=0.024$ \cite{Katz:2005ir}.

The decay $f_2\to\gamma^*\gamma^*$ can be studied from the quadratic action of the electromagnetic field considering the fluctuation of the metric:
\begin{eqnarray}
    S_{f_2\gamma\gamma} &\supset& -A_{f_2} \int d^5x \,\sqrt{g} \, e^{-\phi} \, \mbox{Tr}( F_{MN} F^{MN})
\end{eqnarray}
with $F_{MN}=\partial_M V_N-\partial_N V_M$ and $A_{f_2}=1/4$.
The obtained interaction is of the kind: $1/2\, h_{\mu\nu} T^{\mu\nu}$, where $T^{\mu\nu}$ is the electromagnetic stress-energy tensor.

\begin{figure}[th]
    \centering
    \includegraphics[height=8cm]{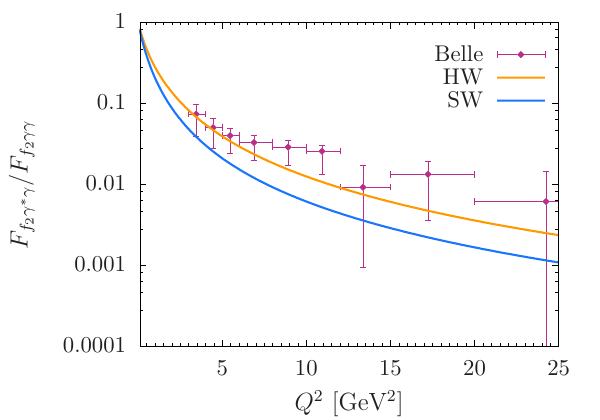}
    \caption{$f_2$ TFF with one real and one virtual photon in the soft-wall (blue curve) and hard-wall (orange curve) models. Belle data are from Ref. \cite{Belle:2015oin}.}
    \label{fig:TFFrv}       
\end{figure}

\begin{figure}[th]
    \centering
    \includegraphics[height=8cm]{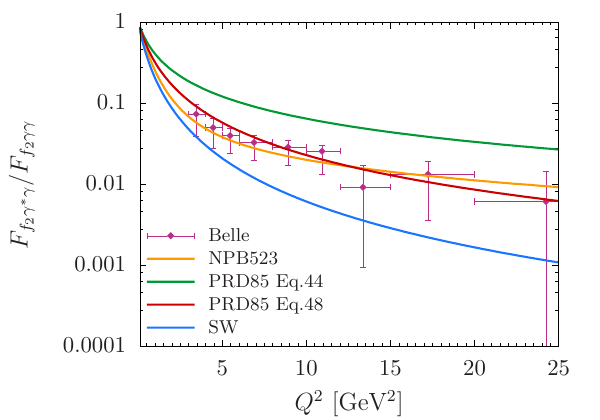}
    \caption{$f_2$ TFF with one real and one virtual photon in the soft-wall model (blue curve), compared with results found in other models: Ref. \cite{Schuler:1997yw} (orange curve), \cite{Pascalutsa:2012pr} (green and red curves). Belle data are from Ref. \cite{Belle:2015oin}.}
    \label{fig:TFFrv-comp}       
\end{figure}

\begin{figure}[th]
    \centering
    \includegraphics[height=8cm]{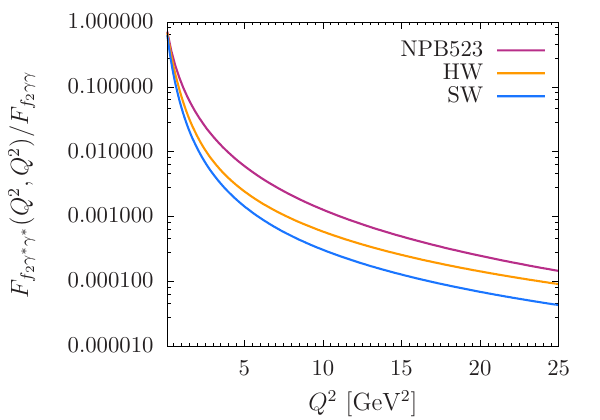}
    \caption{$f_2$ TFF with two virtual photons in the soft-wall model (blue curve), hard-wall (orange curve) model and computed in Ref. \cite{Schuler:1997yw}  (magenta curve).}
    \label{fig:TFFQQ}       
\end{figure}

In the Fourier space, after introducing the source $V_\alpha(q)$ and the BTBP $V(z,q)$ \eqref{eq:VecBTBP} of the vector field, we find:
\begin{eqnarray}\label{eq:ActionTgammagamma}
    S_{f_2\gamma\gamma} &=& 2 A_{f_2} \mbox{Tr}(Q_{em}^2 T^a) \int d^4q_1 \int d^4q_2\int \frac{dz}{z}  e^{-\phi} h^a_{\alpha\beta}(z,q_1+q_2) V_\mu(q_1) V_\nu(q_2)  \nonumber\\
    && (-\eta^{\mu\alpha}\eta^{\nu\beta}\partial_z V(z,q_1) \partial_z V(z,q_2) \nonumber\\
    && + (q_1\cdot q_2 \eta^{\mu\alpha}\eta^{\nu\beta}- q_1^\alpha q_2^\mu \eta^{\nu\beta}+q_1^\alpha q_2^\beta \eta^{\mu\nu}-q_1^\nu q_2^\beta\eta^{\mu\alpha}) V(z,q_1) V(z,q_2) )\,.
\end{eqnarray}  
For a singlet field $T^a=T^{0}$.
To find the amplitude, we derive the action with respect to the three sources of the fields, and write the BTBP of the tensor field as the sum
\begin{equation}
    h(z,q^2) = \sum_n \frac{h_n(z) \sqrt{R_n/k_T}}{-q^2+m_n^2}\,.
\end{equation}
Identifying one transition form factor, the amplitude for $f_2\to \gamma^*\gamma^*$ is:
\begin{eqnarray}\label{eq:f2Amp}
    \mathcal{M}_{\mu\nu\alpha\beta}&=& 4 A_{f_2}  \mbox{Tr}(Q_{em}^2 T^a) \, (q_1\cdot q_2 \,\eta_{\mu\alpha}\eta_{\nu\beta}- q_{1\alpha} \,q_{2\mu} \,\eta_{\nu\beta}+q_{1\alpha} \,q_{2\beta}\, \eta_{\mu\nu}-q_{1\nu}\, q_{2\beta}\, \eta_{\mu\alpha}) \nonumber\\
    && \int_0^\infty \frac{dz}{z}  e^{-\phi} h_0(z) \left(-\frac{1}{q_1\cdot q_2}\partial_z V(z,q_1) \partial_z V(z,q_2)  + V(z,q_1) V(z,q_2)  \right)\,,
\end{eqnarray}
where $h_0(z)$ is the wave function of the ground state, obtained from \eqref{eq:TensorWF} for $n=0$.
The amplitude in \eqref{eq:f2Amp} has the same Lorentz structure given in \cite{Ewerz:2013kda}. 
In \cite{Pauk:2014rta} a different form of the amplitude has been proposed.
From Eq. \eqref{eq:f2Amp} the two-photon transition form factor can be extracted ($q_i^2=-Q_i^2$):
\begin{eqnarray}\label{eq:tensorTFF}
    F_{f_2\gamma^*\gamma^*} (Q_1^2,Q_2^2) &=&  \frac{4}{\sqrt{k_T}} A_{f_2}  \mbox{Tr}(Q_{em}^2 T^a) \, \int_0^\infty \frac{dz}{z}  e^{-\phi} h_{0}^a(z) \nonumber\\
    &&  \left(\frac{-2}{m_0^2+Q_1^2+Q_2^2}\partial_z V(z,Q_1^2) \partial_z V(z,Q_2^2) + V(z,Q_1^2) V(z,Q_2^2)\right)\,.
\end{eqnarray}
Eq. \eqref{eq:tensorTFF} can also be used to compute the TFF in the hard-wall model with some modifications: the upper limit of the integral is $z_0=(322\mbox{ MeV})^{-1}$, the dilaton vanishes, the tensor wave function is \cite{Katz:2005ir}
\begin{equation}
    h_0^{HW}(z) = 3.51 \frac{z^2}{z_0} J_2(m_0 \, z)
\end{equation}
with the tensor mass $m_0=1.23$ GeV, the vector BTBP is \cite{Leutgeb:2019gbz}
\begin{equation}
    V^{HW}(z,Q^2) = \sqrt{Q^2} z \left( K_1(\sqrt{Q^2} z) +I_1(\sqrt{Q^2} z) \frac{K_0(\sqrt{Q^2} z_0)}{I_0(\sqrt{Q^2} z_0)} \right) \,.
\end{equation}
If one photon is real, only the second term in \eqref{eq:tensorTFF} contributes to $F_{f_2\gamma^*\gamma}$, and its behaviour in the soft-wall and hard-wall models is shown in Fig. \ref{fig:TFFrv} together with data 
for the helicity-2 component of $f_2(1270)$  from Belle collaboration \cite{Belle:2015oin}. 
The curve obtained in the hard-wall model has a better agreement with the experimental points, and this is due to the fact that the ground-state mass prediction of the HW is closer to the experimental $f_2(1270)$ mass than the  SW prediction. 
Indeed, if the $f_2(1270)$ mass is used to fix the mass scale $c$ in the soft-wall model, the TFF in the SW and HW models are similar.   
In Fig. \ref{fig:TFFrv-comp} the results from \cite{Schuler:1997yw} and two determinations from \cite{Pascalutsa:2012pr} are also shown.
In Fig. \ref{fig:TFFQQ}  $F_{f_2\gamma^*\gamma^*} (Q^2,Q^2)$ for two photons with equal virtualities is shown.
The $f_2(1270)\to\gamma\gamma$ decay width is $\Gamma_{\gamma\gamma}=2.6\pm 0.5$ keV \cite{ParticleDataGroup:2022pth}, and it is related to the form factor by
\begin{equation}
    \Gamma_{\gamma\gamma}=\frac{\pi\alpha^2}{20} m_0^3 F_{f_2\gamma\gamma}^2\,,
\end{equation}
which fixes $F_{f_2\gamma\gamma}=0.387$ GeV$^{-1}$.
From Eq.~\eqref{eq:tensorTFF}, the form factor for two real photons is $F_{f_2\gamma\gamma}=4 A_{f_2}/\sqrt{k_T} \mbox{Tr}(Q_{em}^2 T^0)/2c$ in the soft-wall model. Using $A_{f_2}=1/4$ we find $F_{f_2\gamma\gamma}=0.986$ GeV$^{-1}$. In the hard-wall model one has $F_{f_2\gamma\gamma}=4 A_{f_2}/\sqrt{k_T} \mbox{Tr}(Q_{em}^2 T^0) (2.084\mbox{ GeV}^{-1})$, hence $A_{f_2}=1/4$ produces $F_{f_2\gamma\gamma}= 1.594$ GeV$^{-1}$.

The results can be extended to non-singlet tensor mesons. 
As previously noticed,   the action \eqref{eq:actionTensor} is used to describe also a generic tensor field dual to the operator $j_{\mu\nu}=\bar q \frac{1}{2} \left(\gamma_\mu i\overleftrightarrow{D}_\nu + \gamma_\nu i\overleftrightarrow{D}_\mu \right)q$, without considering the fluctuations of the metric.
Assuming its coupling to two photons is described by the action \eqref{eq:ActionTgammagamma}, we consider the product $\tilde A_{f_2}=A_{f_2} \mbox{Tr}(Q_{em}^2 T^a)$ as a free parameter and we fix it from $F_{f_2\gamma\gamma}=0.387$ GeV$^{-1}$. We get $\tilde A_{f_2}=0.0267$ in the soft-wall model and $\tilde A_{f_2}=0.0165$ in the hard-wall model. 
We shall use such values of $\tilde A_{f_2}$ to compute the contribution to $a_\mu$ from $f_2(1270)$.

As done for axial-vector mesons, the asymptotic large-$Q^2$ behaviour of the $f_2$ TFF can be analytically obtained from Eq. \eqref{eq:tensorTFF} defining the variables $Q^2=(Q_1^2+Q_2^2)/2$ and $w=(Q_1^2-Q_2^2)/(Q_1^2+Q_2^2)$, noticing that in the soft-wall model $V(z,Q^2)\xrightarrow[Q^2\to \infty]{} z \sqrt{Q^2} K_1(z \sqrt{Q^2})$ \cite{Colangelo:2011xk} and $h_0(z) = f_0 m_0^2/(4\sqrt{k_T}) z^4$.
We find:
\begin{eqnarray}\label{eq:asyTensor}
    F_{f_2\gamma^*\gamma^*} &\xrightarrow[Q^2\to\infty]{}& F_{f_2\gamma\gamma} 2c \frac{f_0 m_0^2}{4 \sqrt{k_T} Q^4} \int_0^\infty dx\, x^5 \nonumber\\
    && [-(1-w^2) K_0(x \sqrt{1+w}) \,K_0(x \sqrt{1-w}) + \sqrt{1-w^2} K_1(x \sqrt{1+w}) \,K_1(x \sqrt{1-w})] \nonumber\\
    &=& \frac{F_{f_2\gamma\gamma}}{4 \sqrt{2 k_T}} \frac{f_0 m_0^3}{Q^4 w^5} \left( 4 w (6 - 5w^2 )+2 (6-7 w^2+w^4) \log\frac{1-w}{1+w}\right) \,.\nonumber\\
\end{eqnarray}
In \cite{Hoferichter:2020lap} this expansion has been computed in perturbative QCD. The form factor multiplying the tensor 
\begin{equation}
    T_1^{\mu\nu\alpha\beta} = 2 q_1\cdot q_2 \eta^{\mu\alpha}\eta^{\nu\beta}-2 q_1^\alpha q_2^\mu \eta^{\nu\beta}+2q_1^\alpha q_2^\beta \eta^{\mu\nu}-2q_1^\nu q_2^\beta\eta^{\mu\alpha}
\end{equation}
at large $Q^2$ is 
\begin{eqnarray}
    \mathcal{F}_1^T(q_1^2,q_2^2) = \frac{4\sum_a C_a F_A^a m_T^3}{Q^4} f_1^T(w)
\end{eqnarray}
with 
\begin{equation}
f_1^T(w) = \frac{5 (1-w^2)}{8 w^6} \left( 15-4 w^2+\frac{3(5-3 w^2)}{2w}  \log\frac{1-w}{1+w}\right)
\end{equation}
having the same $Q^2$ but a different $w$ behaviour than  in Eq. \eqref{eq:asyTensor}.

The results for the TFF can be used to compute the HLbL pole contribution of tensor mesons to muon $g-2$ using Eq. \eqref{eq:amutensorfin} in  Appendix \ref{sec:AppAmuTensor}.
We find that the $f_2(1270)$ contribution to muon $g-2$ is $a^{HLbL}_\mu\sim 0.61\times 10^{-11}$ in the soft-wall model and $a^{HLbL}_\mu\sim 0.63\times 10^{-11}$ in the hard-wall model. 
They agree with the result in \cite{Pauk:2014rta} $a^{HLbL}_\mu = (0.79\pm 0.09)\times 10^{-11}$.
In Ref. \cite{Danilkin:2016hnh} $a^{HLbL}_\mu=(0.50\pm 0.13)\times 10^{-11}$ and $a^{HLbL}_\mu=(0.21\pm 0.05)\times 10^{-11}$ are obtained for $f_2(1270)$ and $f_2(1565)$, respectively.

\section{Conclusions}
Holographic bottom-up models, despite their simplicity, provide a good qualitative description of QCD observables in different sectors, and in some case succeed in quantitative predictions.
This also holds for the two-photon transition form factors of axial-vector and tensor mesons. 
The TFF of tensor mesons in the soft-wall model is smaller than experimental values but the differences are within the experimental errors. 
A better agreement is found between the hard-wall model and experimental data, since the mass prediction better reproduces the measured $f_2(1270)$ mass.
The results for axial-vector and tensor meson contributions to the anomalous magnetic moment of the muon confirm the decreasing hierarchy from pseudoscalar, axial-vector and tensor meson poles.
The values we have found in the soft-wall model are $a_\mu^{HLbL}=36\times 10^{-11}$ for axial-vector mesons and $a_\mu^{HLbL}=0.61\times 10^{-11}$ for tensor mesons. 
The contribution from the pion computed in \cite{Colangelo:2023een} in a model that for pions is identical to the one considered in this paper is $a_\mu^{HLbL}=75.2\times 10^{-11}$ for the ground state and $a_\mu^{HLbL}=1.68\times 10^{-11}$ for the first excited state. 
It would be interesting to extend the present analysis considering the model of Ref.~\cite{Colangelo:2023een} with the strange quark mass and the $U(1)_A$ anomaly. This is left to a future study.

\section*{Acknowledgments}
We thank Luigi Cappiello, Josef Leutgeb, Jonas Mager, Pere Masjuan, Anton Rebhan, Pablo Roig and Pablo Sanchez-Puertas for discussions.
The work is carried out within the INFN project (Iniziativa Specifica) SPIF.
It has been partly funded by the European Union - Next Generation EU through the research grant number P2022Z4P4B ``SOPHYA - Sustainable Optimised PHYsics Algorithms: fundamental physics to build an advanced society'' under the program PRIN 2022 PNRR of the Italian Ministero dell'Universit\`a e Ricerca (MUR).

\appendix
\section{Tensor meson action}\label{sec:AppActionTensor}
Let us compute the Einstein-Hilbert action for the metric
\begin{equation}
    ds^2=g_{MN}dx^M dx^N =\frac{1}{z^2} (\eta_{\mu\nu}+\lambda h_{\mu\nu}) dx^\mu dx^\nu-\frac{1}{z^2} dz^2\,,
\end{equation}
where $h_{\mu\nu}$ is a small fluctuation of the metric and $\lambda$ a parameter. We consider a transverse symmetric tensor, hence  $\partial^\mu h_{\mu\nu}=0$ and $h_{\mu\nu}=h_{\nu\mu}$.
The Einstein-Hilbert action is:
\begin{equation}
    S_{EH} = -2k_T\int d^5x \,\sqrt{g} \, (\mathcal{R}+2\Lambda)\,,
\end{equation}
where $\mathcal{R}$ is the Ricci scalar and $\Lambda$ the cosmological constant.
At $\mathcal{O}(\lambda^0)$ the metric describes a $5d$ AdS space, with $\mathcal{R}^{(0)}=20$ and $\Lambda=-6$.
To  $\mathcal{O}(\lambda^2)$ we find:
\begin{equation}
    g^{\mu\nu} = z^2 (\eta^{\mu\nu} - \lambda h^{\mu\nu} +\lambda^2 h^{\mu\alpha} h_{\alpha}^\nu) + \mathcal{O}(\lambda^3)
\end{equation}
\begin{equation}
    \sqrt{g} = z^{-5} \left(1 + \frac{\lambda}{2} h +\lambda^2 \left(-\frac{1}{4} h^{\mu\nu} h_{\mu\nu} + \frac{1}{8} h^2\right)\right)+ \mathcal{O}(\lambda^3) ,
\end{equation}
where $h^{\mu\nu}=\eta^{\alpha\mu}\eta^{\beta\nu} h_{\alpha\beta}$ and $h=\eta^{\mu\nu} h_{\mu\nu}$,
\begin{eqnarray}
    R_{zz} &=& -\frac{4}{z^2} +\lambda \left(\frac{1}{2z} \partial_z h -\frac{1}{2} \partial_z^2 h \right)+\lambda^2 \left(\frac{1}{2} \partial_z (h^{\alpha\beta} \partial_z h_{\alpha\beta}) -\frac{1}{4} (\partial_z h^{\alpha\beta}) (\partial_z h_{\alpha\beta}) -\frac{1}{2z}h^{\alpha\beta} \partial_z h_{\alpha\beta}\right)\nonumber\\
    &&+ \mathcal{O}(\lambda^3)
\end{eqnarray}
\begin{eqnarray}
    R_{\mu\nu} &=& \frac{4}{z^2}\eta_{\mu\nu}  +\lambda \left( \frac{4}{z^2}h_{\mu\nu} - \frac{3}{2z} \partial_z h_{\mu\nu} +\frac{1}{2} \partial_z^2 h_{\mu\nu} - \frac{1}{2} \Box h_{\mu\nu} - \frac{1}{2} \partial_\mu\partial_\nu h -\frac{1}{2z} \eta_{\mu\nu}\partial_z h \right)+\nonumber\\
    && +\lambda^2 \left(-\frac{1}{2z} h_{\mu\nu} \partial_z h +\frac{1}{4} (\partial_z h) (\partial_z h_{\mu\nu}) +\frac{1}{2z} \eta_{\mu\nu}h^{\alpha\beta} \partial_z h_{\alpha\beta} -\frac{1}{4} (\partial_\beta h) \partial^\beta h_{\mu\nu} +\right.\nonumber\\
    &&\left.- \frac{1}{2} (\partial_z h_{\mu\beta}) \partial_z h_{\nu}^\beta - \frac{1}{4} (\partial_\mu h^{\alpha\beta}) \partial_\nu h_{\alpha\beta} + \frac{1}{2} (\partial_\beta h_\mu^\alpha) \partial^\beta h_{\nu\alpha}\right)+ \mathcal{O}(\lambda^3) \,.
\end{eqnarray}
In $\mathcal{R}$ the $h_{\mu\nu}$ and $h$ decouple. Since we are interested in a spin-2 field, we do not consider  terms depending on $h$:
\begin{eqnarray}
    \mathcal{R} &=& 20+\lambda^2 \left(4 z h^{\alpha\beta} \partial_z h_{\alpha\beta} -\frac{z^2}{4} (\partial_z h^{\alpha\beta}) (\partial_z h_{\alpha\beta}) + \frac{z^2}{4} (\partial_\mu h^{\alpha\beta}) (\partial^\mu h_{\alpha\beta}) - \frac{z^2}{2}  h^{\alpha\beta} \partial_z^2 h_{\alpha\beta} +\right.\nonumber\\
    &&\left. +\frac{z^2}{2} h^{\alpha\beta} \Box h_{\alpha\beta} - \frac{z^2}{2} \partial_z (h^{\alpha\beta} \partial_z h_{\alpha\beta})\right) + \mathcal{O}(\lambda^3)\,.
\end{eqnarray}
Finally, we expand the Lagrangian:
\begin{eqnarray}
    \sqrt{g} (\mathcal{R}+2\Lambda) &=& \frac{8}{z^5}+\lambda^2 \left(\frac{4}{z^4} h^{\alpha\beta} \partial_z h_{\alpha\beta} -\frac{1}{4z^3} (\partial_z h^{\alpha\beta}) (\partial_z h_{\alpha\beta}) + \frac{1}{4z^3} (\partial_\mu h^{\alpha\beta}) (\partial^\mu h_{\alpha\beta})  +\right.\nonumber\\
    &&\left. - \frac{1}{2z^3}  h^{\alpha\beta} \partial_z^2 h_{\alpha\beta}+\frac{1}{2z^3} h^{\alpha\beta} \Box h_{\alpha\beta} - \frac{1}{2z^3} \partial_z (h^{\alpha\beta} \partial_z h_{\alpha\beta})-\frac{2}{z^5} h^{\alpha\beta} h_{\alpha\beta} \right) + \mathcal{O}(\lambda^3)\nonumber\\
    &=& \frac{8}{z^5}+\lambda^2 \left( \partial_z\left( \frac{1}{2z^4} h^{\alpha\beta} h_{\alpha\beta}\right) +\frac{1}{4z^3} (\partial_z h^{\alpha\beta}) (\partial_z h_{\alpha\beta}) + \frac{1}{4z^3}  h^{\alpha\beta} \Box h_{\alpha\beta} +\right.\nonumber\\
    && \left. -  \partial_z \left( \frac{1}{z^3} h^{\alpha\beta} \partial_z h_{\alpha\beta}\right) \right) + \mathcal{O}(\lambda^3)\,.
\end{eqnarray} 
The Gibbons-Hawking-York boundary term is:
\begin{eqnarray}
    S_{GHY} &=& -4k_T \int_\partial d^4 x \sqrt{\gamma} K\,,
\end{eqnarray}
where $\gamma$ is the determinant of the induced metric on the boundary $\partial$ of the spacetime and $K$ the extrinsic curvature. 
At $\mathcal{O}(\lambda^2)$ it reads:
\begin{eqnarray}
    S_{GHY}^{(2)} &=& - 2k_T \int_\partial d^4 x \, \left(-\frac{2}{z^4} h^{\mu\nu} h_{\mu\nu} +\frac{1}{z^3} h^{\mu\nu} \partial_z h_{\mu\nu} \right)\,.
\end{eqnarray} 
The total action at $\mathcal{O}(\lambda^2)$ is then:
\begin{eqnarray}
    S^{(2)} &=& - 2k_T \int d^5x \, \left( \frac{1}{4z^3} (\partial_z h^{\alpha\beta}) (\partial_z h_{\alpha\beta}) + \frac{1}{4z^3}  h^{\alpha\beta} \Box h_{\alpha\beta} \right) + 3k_T \int_\partial d^4 x \, \frac{1}{z^4} h^{\mu\nu} h_{\mu\nu} \,.
\end{eqnarray}
Neglecting the last (boundary) term, the action considered in the soft-wall model in Sec. \ref{Sec:Tensor} is:
\begin{eqnarray}
    S &=& - \frac{k_T}{2} \int d^5x \, e^{-\phi} \left( \frac{1}{z^3} (\partial_z h^{\alpha\beta}) (\partial_z h_{\alpha\beta}) + \frac{1}{z^3}  h^{\alpha\beta} \Box h_{\alpha\beta} \right)   \,.
\end{eqnarray}

\section{$a^{HLbL}_\mu$ from tensor meson poles}\label{sec:AppAmuTensor}
The HLbL contribution to the anomalous magnetic moment of the muon is computed from \cite{Jegerlehner:2009ry}:
\begin{eqnarray}
    a^{HLbL}_\mu&=&-\frac{i e^6}{48 m_\mu} \int \frac{d^4 q_1}{(2\pi)^4} \int \frac{d^4 q_2}{(2\pi)^4}  \frac{1}{q_1^2 q_2^2 (k-q_1-q_2)^2} \frac{1}{(p-q_1)^2-m_\mu^2} \frac{1}{(p-q_1-q_2)^2-m_\mu^2}  \nonumber\\
    && \mbox{Tr}\left(  (\slashed p+m_\mu) [\gamma^\rho,\gamma^\sigma] (\slashed p+m_\mu) \gamma^\mu (\slashed p-\slashed q_1+m_\mu) \gamma^\nu (\slashed p-\slashed q_1-\slashed q_2+m_\mu) \gamma^\lambda \right) \nonumber\\
    &&\left. \frac{\partial}{\partial k^\rho} \Pi_{\mu\nu\lambda\sigma}(q_1,q_2,k-q_1-q_2) \right|_{k\to 0}\,,
\end{eqnarray}
where  $p$ denotes the initial muon momentum, $m_\mu$ is the muon mass, $q_1,q_2,q_3,q_4$ are the momenta of the incoming photons, and $k=-q_4$. 
Let us consider the contribution to $\Pi_{\mu\nu\lambda\sigma}$ from tensor-meson exchange \cite{Pauk:2014rta}:
\begin{eqnarray}
   (ie)^4 \Pi_{\mu\nu\lambda\sigma}^T(q_1,q_2,q_3) &=& \mathcal{M}_{\mu\nu\alpha\alpha'}(q_1,q_2) \frac{i P^{\alpha\alpha'\beta\beta'}(k-q_3)}{(k-q_3)^2-M_T^2} \mathcal{M}_{\lambda\sigma\beta\beta'}(q_3,-k)\nonumber\\
    && + \mathcal{M}_{\mu\sigma\alpha\alpha'}(q_1,-k) \frac{i P^{\alpha\alpha'\beta\beta'}(k-q_1)}{(k-q_1)^2-M_T^2} \mathcal{M}_{\nu\lambda\beta\beta'}(q_2,q_3)\nonumber\\
    && + \mathcal{M}_{\mu\lambda\alpha\alpha'}(q_1,q_3) \frac{i P^{\alpha\alpha'\beta\beta'}(k-q_2)}{(k-q_2)^2-M_T^2} \mathcal{M}_{\nu\sigma\beta\beta'}(q_2,-k)
\end{eqnarray}
with $M_T$ the mass of the tensor meson. We only consider the dominant contribution from helicity $\Lambda=2$ \cite{Pauk:2014rta}, for which the amplitude of the $f_2$ production from two photons is written as \cite{Ewerz:2013kda}  
\begin{equation}\label{eq:tensorAmpNac}
    \mathcal{M}^{\mu\nu\alpha\beta}(q_1,q_2)=e^2 (q_1\cdot q_2 \eta^{\mu\alpha}\eta^{\nu\beta}- q_1^\alpha q_2^\mu \eta^{\nu\beta}+q_1^\alpha q_2^\beta \eta^{\mu\nu}-q_1^\nu q_2^\beta\eta^{\mu\alpha}) F_{T\gamma^*\gamma^*}(q_1^2,q_2^2)\,,
\end{equation}
where $\alpha,\beta$ are $f_2$ indices.
The projection operator for $J=2$ has the form:
\begin{eqnarray}
    P_{\alpha\beta\gamma\delta}(p)&=&\frac{1}{2} ((-g_{\alpha\gamma}+p_\alpha p_\gamma/p^2) (-g_{\beta\delta}+p_\beta p_\delta/p^2)+(-g_{\alpha\delta}+p_\alpha p_\delta/p^2) (-g_{\beta\gamma}+p_\beta p_\gamma/p^2))\nonumber\\
    &&-\frac{1}{3} (-g_{\alpha\beta}+p_\alpha p_\beta/p^2) (-g_{\gamma\delta}+p_\gamma p_\delta/p^2)\,.
\end{eqnarray}

We define the momenta $Q_i=i q_i$ and $P=i p$, obtaining:
\begin{eqnarray}\label{eq:amud4Q}
    a^{HLbL}_\mu&=&\frac{e^6}{48 m_\mu} \int \frac{d^4 Q_1}{(2\pi)^4} \int \frac{d^4 Q_2}{(2\pi)^4}  \frac{1}{Q_1^2 Q_2^2 (Q_1+Q_2)^2} \frac{1}{(P+Q_1)^2+m_\mu^2} \frac{1}{(P-Q_2)^2+m_\mu^2}  \nonumber\\
    && \left( T_1(Q_1,Q_2,P) \frac{F_{T\gamma^*\gamma^*}(Q_1^2,(Q_1+Q_2)^2) F_{T\gamma^*\gamma^*}(Q_2^2,0)}{Q_2^2+M_T^2}\right.\nonumber\\
    && +  T_2(Q_1,Q_2,P) \frac{F_{T\gamma^*\gamma^*}((Q_1+Q_2)^2,Q_2^2) F_{T\gamma^*\gamma^*}(Q_1^2,0)}{Q_1^2+M_T^2}  \nonumber\\
    &&\left. +  T_3(Q_1,Q_2,P) \frac{F_{T\gamma^*\gamma^*}(Q_1^2,Q_2^2) F_{T\gamma^*\gamma^*}((Q_1+Q_2)^2,0)}{(Q_1+Q_2)^2+M_T^2} \right) \,,
\end{eqnarray}
with
\begin{eqnarray}
    T_1 &=& \frac{32 m_\mu}{3 Q_2^2} ((P\cdot Q_2)^2 (-6 (Q_1\cdot Q_2)^2 + 2 Q_1^2 Q_2^2) + Q_2^2 ((8 m_\mu^2 - 4 P\cdot Q_1) (Q_1\cdot Q_2)^2 +\nonumber\\
    &&+ 2 (-2 (P\cdot Q_1)^2 + m_\mu^2 Q_1^2 + 2 P\cdot Q_1 Q_1^2) Q_2^2 + 5 (2 m_\mu^2 - P\cdot Q_1) Q_1\cdot Q_2 Q_2^2) +\nonumber\\
    &&+ P\cdot Q_2 (4 (Q_1\cdot Q_2)^3 + 2 (9 P\cdot Q_1 - 2 Q_1^2) Q_1\cdot Q_2 Q_2^2 + 5 (Q_1\cdot Q_2)^2 Q_2^2 + 10 P\cdot Q_1 Q_2^4))\nonumber\\
\end{eqnarray}
\begin{eqnarray}
    T_3 &=& -\frac{32 m_\mu}{3 (Q_1+Q_2)^2} ((P\cdot Q_2)^2 (4 Q_1^4 + 6 (Q_1\cdot Q_2)^2 +    Q_1^2 (8 Q_1\cdot Q_2 - 2 Q_2^2)) +\nonumber\\
    &&+ (P\cdot Q_1)^2 (6 (Q_1\cdot Q_2)^2 + 8 Q_1\cdot Q_2 Q_2^2 - 
         2 (Q_1^2 - 2 Q_2^2) Q_2^2) -\nonumber\\ 
    &&-  3 P\cdot Q_2 (Q_1^2 + Q_1\cdot Q_2) (Q_1^2 (5 Q_1\cdot Q_2 + 2 Q_2^2) + 
         Q_1\cdot Q_2 (8 Q_1\cdot Q_2 + 5 Q_2^2)) +\nonumber\\ 
    &&+  2 m_\mu^2 (Q_1^4 (5 Q_1\cdot Q_2 + 4 Q_2^2) + 
         2 Q_1^2 (8 (Q_1\cdot Q_2)^2 + 9 Q_1\cdot Q_2 Q_2^2 + 
            2 Q_2^4) + \nonumber\\
    &&+ Q_1\cdot Q_2 (12 (Q_1\cdot Q_2)^2 + 16 Q_1\cdot Q_2 Q_2^2 + 
            5 Q_2^4)) + \nonumber\\
    &&+  P\cdot Q_1 (2 P\cdot Q_2 (5 Q_1^4 + 18 (Q_1\cdot Q_2)^2 + 
            16 Q_1\cdot Q_2 Q_2^2 + 5 Q_2^4 + 
            4 Q_1^2 (4 Q_1\cdot Q_2 + Q_2^2)) +\nonumber\\ 
    && + 3 (Q_1\cdot Q_2 + Q_2^2) (Q_1^2 (5 Q_1\cdot Q_2 + 2 Q_2^2) + 
            Q_1\cdot Q_2 (8 Q_1\cdot Q_2 + 5 Q_2^2))))\,.
\end{eqnarray}
The second term in \eqref{eq:amud4Q} coincides with the first one after changing variables $Q_1 \to -Q_2$ and $Q_2 \to -Q_1$.
To eliminate the dependence on the direction of the muon momentum $P$, we average over all spatial directions of $P$
\begin{equation}
    a_\mu^{HLbL} = \frac{1}{2\pi^2} \int d\Omega(\hat P) \,\, a_\mu^{HLbL}\,.
\end{equation}
The integrals can be done analytically expressing the propagators in terms of Gegenbauer polynomials \cite{Jegerlehner:2009ry}.
Defining $t=\cos\theta$, with $\theta$ the angle between the four vectors $Q_1$ and $Q_2$, and $Q_1=|Q_1|$ and $Q_2=|Q_2|$, we obtain
\begin{eqnarray}\label{eq:amutensorfin}
    a_\mu^{HLbL}&=&\frac{\alpha^3}{24 \pi^2 m_\mu} \int_0^\infty dQ_1 \int_0^\infty d Q_2 \int_{-1}^1 dt \, \sqrt{1-t^2} \,\frac{Q_1 Q_2}{Q_3^2}  \nonumber\\
    && \left( I_1(Q_1,Q_2,t) \frac{F_{T\gamma^*\gamma^*}(Q_1^2,Q_3^2) F_{T\gamma^*\gamma^*}(Q_2^2,0)}{Q_2^2+M_T^2}\right)\nonumber\\
    && \left.+ I_2(Q_1,Q_2,t) \frac{F_{T\gamma^*\gamma^*}(Q_1^2,Q_2^2) F_{T\gamma^*\gamma^*}(Q_3^2,0)}{Q_3^2+M_T^2}\right)\,,
\end{eqnarray}
with
\begin{eqnarray}
    I_1(Q_1,Q_2,t)&=& \frac{ 8 Q_2}{3 m_\mu} (-10 Q_2^3 - 6 Q_1 Q_2^2 t - 4 m_\mu^2 (5 Q_2 + 9 Q_1 t) + 4 Q_1^2 Q_2 (-5 + 3 t^2) +\nonumber\\
    &&+ Q_1^3 t (-27 + 11 t^2) - 2 Q_1^2 (2 Q_2 + 7 Q_1 t) (-2 + t^2) R_1 + Q_1^3 t (-1 + 3 t^2) R_1^2 +\nonumber\\
    &&+ 2 Q_2 R_2 (5 Q_2^2 + 2 Q_1 Q_2 t + Q_1^2 (6 - 4 t^2) + Q_1 Q_2 t R_2) + \nonumber\\
    &&+8 Q_1 Q_2 (2 m_\mu^2 (Q_1 + 5 Q_2 t + 4 Q_1 t^2) + Q_1 (Q_2^2 (-2 + t^2) + 2 Q_1 Q_2 t (-2 + t^2) +\nonumber\\
    && Q_1^2 (-3 + 2 t^2))) X(Q_1, Q_2, t)) 
\end{eqnarray}
\begin{eqnarray}
    I_2(Q_1,Q_2,t)&=& \frac{4}{3 m_\mu Q_3^2 } (2 (5 Q_1^6 + 5 Q_2^6 + 32 Q_1^5 Q_2 t + 32 Q_1 Q_2^5 t + Q_1^3 Q_2^3 t (65 + 51 t^2) + \nonumber\\
    && +Q_1^4 Q_2^2 (14 + 71 t^2) + 
         Q_1^2 Q_2^4 (14 + 71 t^2) + \nonumber\\
    &&+  2 m_\mu^2 (5 Q_1^4 + 5 Q_2^4 + 16 Q_1^3 Q_2 t + 16 Q_1 Q_2^3 t + 
            2 Q_1^2 Q_2^2 (2 + 9 t^2))) - \nonumber\\
    &&-  2 Q_1^2 (5 Q_1^4 + 33 Q_1^3 Q_2 t + 6 Q_2^4 (1 + 2 t^2) + 
         Q_1 Q_2^3 t (32 + 27 t^2) + Q_1^2 Q_2^2 (8 + 61 t^2)) R_1 +\nonumber\\ 
    && +  Q_1^3 Q_2 t (2 Q_1^2 + 4 Q_1 Q_2 t + Q_2^2 (-1 + 3 t^2)) R_1^2 + \nonumber\\
    &&+  Q_2 (Q_2 R_2 (-2 (5 Q_2^4 + 33 Q_1 Q_2^3 t + 6 Q_1^4 (1 + 2 t^2) + 
               Q_1^3 Q_2 t (32 + 27 t^2) + \nonumber\\
    && + Q_1^2 Q_2^2 (8 + 61 t^2)) + 
            Q_1 Q_2 t (2 Q_2^2 + 4 Q_1 Q_2 t + Q_1^2 (-1 + 3 t^2)) R_2) + \nonumber\\
    &&+     8 Q_1 (Q_1^2 + Q_2^2 + 
            2 Q_1 Q_2 t) (-2 m_\mu^2 (4 Q_1 Q_2 + 5 (Q_1^2 + Q_2^2) t + 
               6 Q_1 Q_2 t^2) +\nonumber\\ 
    &&+        3 Q_1 Q_2 (Q_1^2 + Q_2^2 + 6 Q_1 Q_2 t + 2 (Q_1^2 + Q_2^2) t^2)) X(Q_1, Q_2, t))) 
\end{eqnarray}
and $Q_3^2=Q_1^2+2 Q_1 Q_2 t+Q_2^2$, $X(Q_1,Q_2,t)=\frac{1}{Q_1 Q_2 \sqrt{1-t^2}} \arctan\left(\frac{z\sqrt{1-t^2}}{1-zt}\right) $, $z=\frac{Q_1 Q_2}{4m_\mu^2}(1-R_{1})(1-R_{2})$, $R_{i}=\sqrt{1+4m_\mu^2/Q_i^2}$.

\bibliographystyle{JHEP}
\bibliography{text.bib}
\end{document}